\documentclass[useAMS, usenatbib]{mn2e}
 \usepackage{graphicx,amsmath}

\begin{document}

 \title[Galactic Parameters from Masers]
  {Galactic Parameters from Masers with Trigonometric Parallaxes}
 \author[V.V. Bobylev \& A.T. Bajkova]
  {Vadim V.~Bobylev and Anisa T.~Bajkova\\
  Central Astronomical Observatory at Pulkovo of RAS,
 Pulkovskoye Chaussee 65/1, 196140, Saint-Petersburg, Russia.\\
 E-mail: vbobylev@gao.spb.ru, anisabajkova@rambler.ru}

\date{Accepted 2010 January 00. Received 2010 January 00; in original form 2010 January 00}

\pagerange{\pageref{firstpage}--\pageref{lastpage}} \pubyear{2010}
\def\LaTeX{L\kern-.36em\raise.3ex\hbox{a}\kern-.15em
    T\kern-.1667em\lower.7ex\hbox{E}\kern-.125emX}
\newtheorem{theorem}{Theorem}[section]
\label{firstpage}
\maketitle

\begin{abstract}

Spatial velocities of all currently known 28 masers having
trigonometric parallaxes, proper motion and line-of-site
velocities are reanalyzed using Bottlinger's equations. These
masers are associated with 25 active star-forming regions and are
located in the range of galactocentric distances $3<R<14$~kpc. To
determine the Galactic rotation parameters, we used the first
three Taylor expansion terms of angular rotation velocity $\Omega$
at the galactocentric distance of the Sun $R_0=8$~kpc. We obtained
the following solutions:
 $\Omega_0  =-31.0\pm1.2\;\mathrm{km\,s}^{-1}\mathrm{kpc}^{-1}$,
 ${\Omega_0^{\prime}}= 4.46\pm0.21\;\mathrm{km\,s}^{-1}\,\mathrm{kpc}^{-2}$,
 $\Omega_0^{\prime\prime}=-0.876\pm0.067\;\mathrm{km\,s}^{-1}\,\mathrm{kpc}^{-3}$,
Oort constants:
 $A= 17.8\pm0.8$ km s$^{-1}$ kpc$^{-1}$,
 $B=-13.2\pm1.5$ km s$^{-1}$ kpc$^{-1}$ and circular velocity of the Solar neighborhood rotation
 $V_0=248\pm14\;\mathrm{km\,s}^{-1}$.
Fourier analysis of galactocentric radial velocities of masers
$V_R$ allowed us to estimate the wavelength
$\lambda=2.0\pm0.2$~kpc  and peak velocity $f_R=6.5\pm2$~km
s$^{-1}$ of periodic perturbations from the density wave and
velocity of the perturbations $4\pm1\;\mathrm{km\,s}^{-1}$ near
the location of the Sun. Phase of the Sun in the density wave is
estimated as $\chi_\odot\approx-130^o\pm10^o$. Taking into account
perturbations evoked by spiral density wave we obtained the
following non-perturbed components of the peculiar Solar velocity
with respect to the local standard of rest (LSR)
$(U_\odot,V_\odot,W_\odot)_\mathrm{LSR}=(5.5,11,8.5)\pm(2.2,1.7,1.2)\;\mathrm{km\,
s}^{-1}$.
\end{abstract}

\begin{keywords}
Masers -- SFRs -- Spiral Arms: Rotation Curve -- Galaxy (Milky
Way).
\end{keywords}

\section{INTRODUCTION}

Study of Galactic parameters, especially of the Galactic rotation
curve, is of great importance in solving a number of problems,
such as estimating the mass of the Galaxy, determining the
distribution of matter, estimating the hidden mass, studying
dynamics and structure of the Galaxy and its subsystems, etc.

For this purpose masers having trigonometric parallaxes,
line-of-sight velocities and proper motions (Reid et al., 2009)
are of great interest. Currently measurements of parallaxes with
mean error of several percents are fulfilled with radio
interferometers for three tens of masers in the various active
high-mass star-forming regions. Because these masers are
associated with very young ($<10^5$~yr) OB stars, their kinematics
reflects the properties of the youngest part of the Galactic disk.

Analysis of motions of 18 masers was made by a number of authors
(Reid et al., 2009; Baba et al., 2009; Bovy et al., 2009;
McMillan~\&~Binney, 2010). For the first time Reid et al. (2009)
suggested that high-mass star-forming regions significantly
($\approx 15$ km s$^{-1}$) lag circular rotation. McMillan \&
Binney (2010) suggest that much of this lag could be accounted for
by increasing the value of the Solar motion in the direction of
galactic rotation from 5 to 11 km s$^{-1}$.

The question of particular interest is the peculiar velocity of
the Sun with respect to the LSR,
$(U_\odot,V_\odot,W_\odot)_\mathrm{LSR}$. On the basis of
Str\"omberg's relation, Dehnen~\&~Binney~(1998) determined the
vector
$(U_\odot,V_\odot,W_\odot)_\mathrm{LSR}=(10.0,5.3,7.2)\pm(0.4,0.6,0.4)\;\mathrm{km\,
s}^{-1}$ using proper motions of $\approx$12000 main sequence
stars from the Hipparcos catalogue. These results were confirmed
by Aumer~\&~Binney~(2009):
$(U_\odot,V_\odot,W_\odot)_\mathrm{LSR}=(10.0,5.3,7.1)\pm(0.3,0.5,0.3)\;\mathrm{km\,
s}^{-1}$, who applied the same method to proper motions from the
revised version of the Hipparcos catalogue (van Leeuwen, 2007). In
the work of Sch\"onrich et al.~(2010), where gradient of
metallicity of stars in the Galactic disk was taken into account,
this velocity is different:
$(U_\odot,V_\odot,W_\odot)_\mathrm{LSR}=(11.1,12.2,7.3)\pm(0.7,0.5,0.4)\;\mathrm{km\,
s}^{-1}$. Using another approach, Francis~\&~Anderson~(2009)
suggested that this velocity is:
$(U_\odot,V_\odot,W_\odot)_\mathrm{LSR}=(7.5,13.5.2,6.8)\pm(1.0,0.3,0.1)\;\mathrm{km\,
s}^{-1}$.

In the present work, we are trying to establish relationship
between motions of all currently known masers having parallaxes,
proper motions and line-of-sight velocities, and parameters of the
Galactic spiral density waves (Lin~\&~Shu, 1964), and to estimate
components of the peculiar velocity of the Sun with respect to the
LSR. This goal is achieved by determining parameters of the
Galactic rotation curve, as well as other kinematic parameters, by
means of Bottlinger's equations. Fourier analysis of periodic
deviations of observational velocities of masers from the Galactic
rotation curve is aimed to obtain some estimates of spiral density
wave parameters.

\section{DATA}

Input data on methanol masers with trigonometric parallaxes and
proper motions from VLBI measurements are taken from the papers of
Reid et al. (2009), Rygl et al. (2010). We added also H$_2$O
masers SVS~13 in NGC~1333 (Hirota et al., 2008a),
IRAS~22196$+$6336 in Lynds 1204G (Hirota et al., 2008b) and
G14.33--0.64 (Sato et al., 2010), with parallaxes and proper
motions measured in the framework of VERA (VLBI Exploration of
Radio Astrometry) program. Note that, for maser NGC~281--W listed
in Reid et al.~(2009), we used observational data obtained by Rygl
et al. (2010) with the quoted uncertainty for 6.7 GHz masers
smaller than the Sato et al. (2008) 22 GHz masers. For the nearest
star-forming region in Orion, along with the data on maser
Orion--KL (Hirota et al., 2007), we use independent VLBA
observations of radio star GMR~A (Sandstrom et al., 2007) with
line-of-sight velocity with respect to the LSR
$(V_r)_\mathrm{LSR}=-4\pm5$~km s$^{-1}$ (Bower et al., 2003). It
is the Bower's et al. (2003) opinion that the GMR~A source is a
young star of T~Tau type (WTTS). We use also a very young object a
supergiant S~Per (Asaki et al.,~2007) with line-of-sight velocity
$V_r=-39.7\pm0.7\;\mathrm{km\,s}^{-1}$ from Famaey et al.~(2005).

All 28 objects being studied are listed in Table~1. In Fig.~1, one
can see only 25 points because $X,Y$ coordinates of some objects
are very close to each other. For example, three masers Cep~A,
IRAS~22198 and L~1206 form practically a single point in the $XY$
plane.

\begin{figure}
\begin{center}
\includegraphics[width = 75mm]   {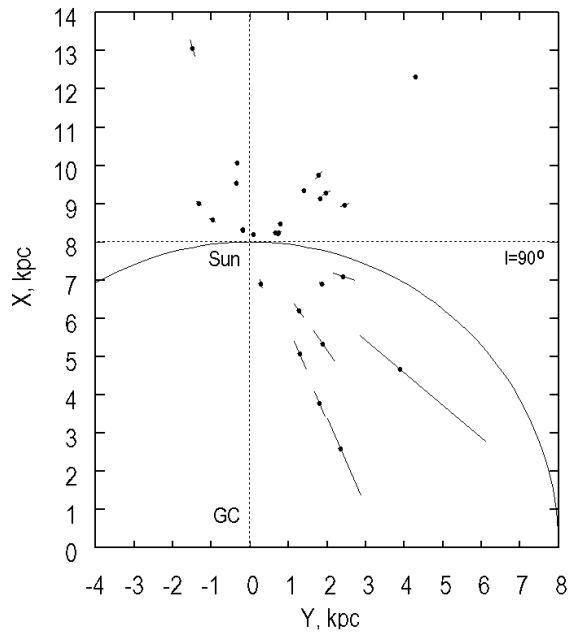}
\caption{Coordinates of masers in the $XY$ Galactic plane.
Locations of the Sun and Galactic Centre (GC) are indicated. The
circle of 8~kpc radius is drawn.}
\end{center}
\end{figure}

\begin{table*}
\caption{Data on masers. Velocities $U^1,V^1,W^1$ are residual
velocities with Galactic rotation excluded using parameters (16).}
 \label{t1}
 \begin{tabular}{@{}lrrrrrrc}
 \hline
 Source          & R, kpc & ~~~~~$U^1,$ km s$^{-1}$ & $V^1,$ km s$^{-1}$ & $W^1,$ km s$^{-1}$ &
                $\Delta V_{\theta},$ km s$^{-1}$ & $V_R,$ km s$^{-1}$ & Ref
\\\hline
L 1287            &  8.52 & $ -2.9\pm1.7$ & $-20.7\pm2.6$ & $-10.2\pm2.5$ & $ -5.7\pm2.6$ & $ -5.6\pm1.7$ & (2) \\
IRAS 00420$+$5530 &  9.33 & $  3.0\pm2.7$ & $-19.6\pm4.2$ & $ -1.5\pm0.8$ & $ -2.9\pm4.2$ & $-11.8\pm2.8$ & (1) \\
NGC 281--W        &  9.50 & $ -4.2\pm2.6$ & $ -2.3\pm2.8$ & $-16.0\pm1.7$ & $ 12.8\pm2.8$ & $ -1.2\pm2.6$ & (2) \\
W3 (OH)           &  9.46 & $  4.8\pm2.1$ & $-17.6\pm2.2$ & $ -6.2\pm0.2$ & $ -1.2\pm2.2$ & $-13.1\pm2.1$ & (1) \\
WB 89--437        & 13.05 & $  0.2\pm2.1$ & $-17.1\pm2.1$ & $  2.5\pm3.8$ & $  0.2\pm2.1$ & $ -8.6\pm2.1$ & (1) \\
NGC 1333 (f1)     &  8.21 & $-23.6\pm4.4$ & $-13.5\pm2.3$ & $  0.0\pm2.5$ & $  0.9\pm2.3$ & $ 15.6\pm4.4$ & (3) \\
NGC 1333 (f2)     &  8.21 & $-21.1\pm4.5$ & $  5.5\pm1.8$ & $ 10.0\pm3.5$ & $ 20.0\pm1.8$ & $ 13.3\pm4.5$ & (3) \\
Ori GMR A         &  8.32 & $ -4.7\pm4.1$ & $ -8.8\pm2.3$ & $ -3.1\pm1.7$ & $  5.6\pm2.3$ & $ -3.4\pm4.1$ & (4) \\
Ori KL            &  8.34 & $-17.9\pm4.2$ & $-14.4\pm2.4$ & $ -3.8\pm3.0$ & $  0.3\pm2.4$ & $  9.9\pm4.2$ & (1) \\
S252 A            & 10.08 & $-12.6\pm3.0$ & $-11.1\pm0.6$ & $ -9.2\pm0.3$ & $  3.5\pm0.6$ & $  4.5\pm3.0$ & (1) \\
S255          &  9.56 & $ -7.7\pm3.6$ & $ -0.8\pm9.5$ & $ -4.0\pm10$  & $ 13.7\pm9.5$ & $ -0.8\pm3.6$ & (2) \\
S269          & 13.16 & $ -3.5\pm2.9$ & $-16.2\pm1.2$ & $-11.7\pm0.9$ & $ -2.2\pm1.2$ & $ -4.3\pm2.9$ & (1) \\
VY CMa        &  8.64 & $ -6.1\pm2.9$ & $-16.9\pm3.0$ & $-13.6\pm2.5$ & $ -2.7\pm3.0$ & $ -1.6\pm2.9$ & (1) \\
G232.62$+$0.9 &  9.12 & $ -7.8\pm3.2$ & $-10.5\pm3.1$ & $ -6.4\pm2.2$ & $  4.0\pm3.1$ & $ -0.9\pm3.2$ & (1) \\
G14.33$-$0.64 &  6.92 & $    0\pm10$  & $  -10\pm14$  & $-10.8\pm4.5$ & $    5\pm14$  & $   -8\pm10$  & (1) \\
G23.66$-$0.13 &  5.23 & $ 21.6\pm3.7$ & $ -6.7\pm5.7$ & $ -3.1\pm0.4$ & $ 14.8\pm5.6$ & $-26.8\pm3.8$ & (1) \\
G23.01$-$0.41 &  4.18 & $  1.0\pm4.7$ & $-20.7\pm9.1$ & $ -8.5\pm2.3$ & $ -1.8\pm8.4$ & $-10.8\pm5.8$ & (1) \\
G23.43$-$0.18 &  3.50 & $  -15\pm10$  & $    9\pm22$  & $ -5.3\pm1.8$ & $   13\pm18$  & $ 20\pm16$    & (1) \\
G35.20$-$0.74 &  6.34 & $-21.7\pm3.3$ & $-20.1\pm3.6$ & $-15.7\pm1.5$ & $ -8.2\pm3.6$ & $ 12.2\pm3.3$ & (1) \\
W48           &  5.65 & $-31.1\pm5.7$ & $-16.5\pm7.5$ & $-16.5\pm2.4$ & $ -9.6\pm7.3$ & $ 21.1\pm5.9$ & (1) \\
W51           &  6.08 & $  -24\pm41$  & $  -10\pm35$  & $-10.4\pm3.7$ & $   -7\pm37$  & $   15\pm38$  & (1) \\
V645          &  7.16 & $-14.4\pm2.6$ & $-13.6\pm2.9$ & $-11.4\pm0.5$ & $ -0.8\pm2.9$ & $  6.3\pm2.6$ & (1) \\
Onsala1       &  7.50 & $   -9\pm13$  & $-18.3\pm5.6$ & $ -2.9\pm6.0$ & $ -4.1\pm6.8$ & $    0\pm12$  & (2) \\
IRAS 22198    &  8.26 & $ -4.2\pm3.7$ & $-23.4\pm4.9$ & $  3.0\pm2.1$ & $ -8.6\pm4.9$ & $ -4.6\pm3.7$ & (3) \\
L 1206        &  8.27 & $-12.8\pm4.4$ & $-19.4\pm3.2$ & $ -7.2\pm5.7$ & $ -5.3\pm3.2$ & $  4.4\pm4.3$ & (2) \\
Cep A         &  8.26 & $ -8.9\pm3.8$ & $-16.6\pm4.9$ & $-12.6\pm1.2$ & $ -2.2\pm4.8$ & $  0.7\pm3.8$ & (1) \\
NGC 7538      &  9.30 & $ -2.1\pm2.1$ & $-32.7\pm2.9$ & $-18.1\pm1.0$ & $-16.0\pm2.8$ & $-10.5\pm2.2$ & (1) \\
S Per         &  9.30 & $-16.1\pm1.2$ & $-16.9\pm1.2$ & $-17.2\pm3.1$ & $ -3.9\pm1.1$ & $  7.5\pm1.1$ & (5) \\

\hline
 \end{tabular}
 \medskip

 {
 Notes. Initial data are taken from:
 (1) -- Reid et al. (2009);  (2) -- Rygl et al. (2010); (3) -- Hirota et al. (2008a);
  (4) -- Sandstrom et al. (2007);
 (5) -- Asaki et al. (2007).}
\end{table*}

Fig.~1 is drawn in galactocentric Cartesian coordinates $X$ and
$Y$. As one can see from this picture, only one maser has inaccurately measured parallax.
That is the maser W51 (in the first Galactic
quadrant), with relative error of parallax $e_\pi/\pi=0.36$.
In fact, only this maser requires calculation of
upper and lower bounds on its distance. But, because this maser
lies in the horizontal part of the Galactic rotation curve, hereafter we use
an average estimate of the heliocentric distance $r\pm
\sigma_r$ or galactocentric distance $R\pm \sigma_R$. The same
distance estimates are used also for other masers.

We reduce line-of-sight velocities of masers with respect to the
LSR to the heliocentric coordinate system using the
standard Solar motion velocity: $(\alpha,
\delta)_{1900}=(270^\circ, +30^\circ)$, $V=20\;\mathrm{km\,s}^{-1}$, hence
$(U_\odot,V_\odot,W_\odot)_\mathrm{LSR}=(10.3,15.3,7.7)\;\mathrm{km\,
s}^{-1}$ (Reid et al., 2009).

It is very important for the purposes of our study to know the
accurate value of $R_0$. A review of $R_0$ estimates made by a
number of authors can be found in the paper by Avedisova~(2005),
where the average value is derived as $R_0=7.80\pm0.33$~kpc. Note
that the most reliable values of this parameter were obtained only
recently. Analysis of motion of the star S2 around the black hole
in the Galactic Centre gave $R_0=8.4\pm0.4$~kpc (Ghez et al.,
2008). This value is in a good agreement with observations of
orbits of 28 stars near the Galactic Centre during 16 yr:
$R_0=8.31\pm0.33$~kpc (Gillessen et al., 2009).

In general, modern estimations show that the most probable value
of $R_0$ is close to $8.0$~kpc, and error of its determination is
close to $0.3$~kpc. For the sake of reliability, we'll consider that the value
of $R_0$ lies in the $7.5-8.5$~kpc range.

\section{METHODS}
\subsection{Determining Rotation Curve Parameters}

Method used here is based on the well-known Bottlinger's formulas
(Ogorodnikov, 1958),
where angular velocity of Galactic rotation is
expanded in a series to $n$-th order terms in
$r/R_0$:
$$
\displaylines{
 V_r=-u_\odot\cos b\cos l
     -v_\odot\cos b\sin l
     -w_\odot\sin b -
 \hfill\llap(1)\cr
 -R_0\sin l \cos b[ (R-R_0) \Omega^1_0/1!
 +\ldots+(R-R_0)^n  \Omega^n_0/n!],
  \hfill\cr
 V_l= u_\odot\sin l - v_\odot\cos l + r \Omega_0 \cos b -
 \hfill\llap(2)\cr
-(R_0\cos l - r\cos b) [(R-R_0)\Omega^1_0/1! +
  \hfill\cr
 +\ldots+(R-R_0)^n\Omega^n_0/n!],
  \hfill\cr
 V_b=u_\odot\cos l \sin b
         +v_\odot\sin l \sin b
         -w_\odot\cos b +
 \hfill\llap(3)\cr
 +R_0\sin l \sin b[(R-R_0)\Omega^1_0/1!
 +\ldots+(R-R_0)^n\Omega^n_0/n!],\hfill
 }
$$
where $V_r$ is heliocentric line-of-sight velocity; $V_l =
4.74r\mu_l \cos b$ and $V_b = 4.74r\mu_b$ are proper motion
velocity components in the $l$ and $b$ directions, respectively
(the coefficient $4.74$ is the quotient of the number of
kilometers in astronomical unit by the number of seconds in a
tropical year); $r=1/\pi$ is the heliocentric distance of an
object; proper motion components $\mu_l \cos b$ and $\mu_b$ are in
$\mathrm{mas\, yr}^{-1}$, line-of-sight velocity $V_r$ is in
$\mathrm{km\,s}^{-1}$; $u_\odot,v_\odot,w_\odot$ are Solar
velocity components with respect to the mean group velocity under
consideration; $R_0$ is the galactocentric distance of the Sun;
$R$ is the galactocentric distance of an object; $R$, $R_0$ and
$r$ are in kpc. Velocity $U$ directed towards the Galactic Centre,
 $V$ along the Galactic rotation, and
 $W$ towards the Northern Galactic pole.
The quantity $\Omega_0$ is the Galactic angular rotational
velocity at distance $R_0$, parameters $\Omega^1_0, \ldots,
\Omega^n_0$ are derivatives of the angular velocity from the first
to the $n$-th order, respectively. The distance $R$ can be
calculated using the expression
$$
\displaylines{\hfill
 R^2=r^2\cos^2 b-2R_0 r\cos b\cos l+R^2_0.\hfill\llap(4)
 }
$$
The system of conditional equations (1)--(3) contains $n+4$
unknowns: $u_\odot$,$v_\odot$,$w_\odot$,
 $\Omega_0,\Omega^1_0,\ldots,\Omega^n_0$,
which can be determined by the least-squares method. The system of
equations (1)--(3) is solved with weights of the form
$$
\displaylines{\hfill
 P_{(r, l, b)}=S_0/\sqrt{S_0^2+\sigma^2_{V_{r, l, b} }},\hfill\llap(5)
 }
$$
where $P_r, P_l$ and $P_b$ are the weights of equation for
$V_r$, $V_l$ and $V_b$ correspondingly, and $S_0=8\;\mathrm{km\,s}^{-1}$ is the ``cosmic'' dispersion
averaged over all observations. Errors
$\sigma_{V_l}$ and $\sigma_{V_b}$ of velocities $V_l$ and
$V_b$ can be calculated using the formulas
$$
\displaylines{\hfill
 \sigma_{(V_l, V_b)} = 4.74 r
 \sqrt{\mu^2_{l, b}\Biggl({\sigma_\pi\over \pi}\Biggr)^2+\sigma^2_{\mu_{l, b}}}.\hfill\llap(6)
 }
$$

\subsection{Determining Projections $V_{\theta}$ and $V_R$}

Components of spatial velocities $U,V,W$ of masers are
determined from observed radial (line-of-sight) and tangential
velocities $V_r,V_l,V_b$ in the following way:
$$
\displaylines{\hfill
 U=V_r\cos l\cos b-V_l\sin l-V_b\cos l\sin b,\hfill\cr\hfill
 V=V_r\sin l\cos b+V_l\cos l-V_b\sin l\sin b,\hfill\llap(7)\cr\hfill
 W=V_r\sin b                +V_b\cos b.\hfill
 }
$$
Then we find two projections of these velocities: $V_R$, directed
radially from the Galactic Centre towards an object, and
$V_{\theta}$, orthogonal to $V_R$ and directed towards Galactic
rotation:
$$
\displaylines{\hfill
  V_{\theta}= U\sin \theta+(V_0+V)\cos \theta, \hfill\llap(8)\cr\hfill
  V_R=-U\cos \theta+(V_0+V)\sin \theta, \hfill\llap(9)
}
$$
where $V_0=|R_0\Omega_0|$, and position angle $\theta$ is
determined as $\tan\theta=y/(R_0-x)$, where $x,y$ are Galactic
Cartesian coordinates of an object. In addition, it is assumed
that velocities $U$ and  $V$ are free from the Solar velocity with
respect to mean group velocity $(u_\odot,v_\odot,w_\odot)$
obtained from equations (1)--(3). Errors of projections $V_R$ and
$V_{\theta}$ are estimated from the expressions
$$
\displaylines{\hfill
  \sigma^2(V_{\theta})= \sigma^2_U\sin^2 \theta+\sigma^2_V\cos^2 \theta, \hfill\llap(10)\cr\hfill
  \sigma^2(V_R)= \sigma^2_U\cos^2 \theta+\sigma^2_V\sin^2 \theta, \hfill\llap(11)
}
$$
where errors of spatial velocities $U$ and $V$ are denoted by
$\sigma_U$ and $\sigma_V$, respectively.

We are interested also in heliocentric spatial velocities which
are free from Galactic rotation with parameters (16) but not from
Solar velocity. We denote such velocities by $U^1,V^1,W^1$ (see
Table~1 and Fig.~7). These velocities must satisfy the following
equations:
$$
\displaylines{\hfill
 {\overline U}^1= -u_\odot,\quad
 {\overline V}^1= -v_\odot,\quad
 {\overline W}^1= -w_\odot.\hfill\llap(12)
 }
$$

\begin{figure}
\begin{center}
\includegraphics[width = 75mm]   {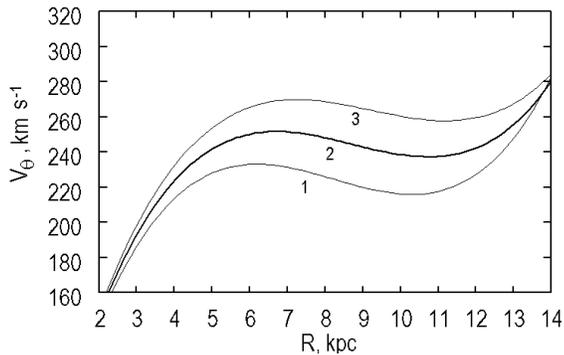}
\caption{Galactic rotation curves
 $V_{\theta}(R)$ for adopted $R_0=7.5,8.0,8.5$~kpc, lines 1,2 and 3.}
\end{center}
\end{figure}

\subsection{Fourier Analysis of Velocities}

At the next step, we consider deviations from circular velocities
$\Delta V_{\theta}$ and galactocentric radial velocities $V_R$.

Spectral analysis consists in applying direct Fourier transform to
the sequence of residual velocities following way:
$$
\displaylines{\hfill
 \overline{V}(\lambda_k)=
 \frac{1}{M}\sum_i^M
 V^i\exp\left(-j\frac{2\pi}{\lambda_k} R_i \right),\hfill\llap(13)
}
$$
where $\overline{V}(\lambda_k)$ is the $k$-th harmonic of Fourier
transform, $M$ is the number of measurements of velocities $V^i$
with coordinates $R_i$,
 $i=1,2,\ldots,M$, and $\lambda_k$ is wavelength in kpc; the latter is equal
to $D/k$, where $D$ is the period of the original sequence in kpc.
In density wave theory, $\lambda$ denotes the distance between
adjacent spiral arms along the Galactic radius vector.

Understanding the results of spectral analysis is easy in the
framework of linear theory of density waves. According to
this approach (Burton~\&~Bania, 1974; Byl~\&~Ovenden, 1978),
$$
\displaylines{\hfill
       V_R =-f_R \cos \chi,\hfill\llap(14)\cr\hfill
 \Delta V_{\theta}= f_\theta \sin\chi,~\hfill\llap(15)
 }
 $$
where $\chi=m[\cot(i)\ln(R/R_0)-\theta]+\chi_\odot$  is phase of
the spiral wave ($m$ is number of spiral arms, $i$ is pitch angle,
$\chi_\odot$ is the phase of the Sun in the spiral wave (Rohlfs,
1977)); $f_R$ and $f_\theta$ are amplitudes of radial and
tangential components of the perturbed velocities which, for
convenience, are always considered positive.

For the case when perturbations $V_R$ and $\Delta V_\theta$ are
considered as periodic functions of logarithm of $R$, the
following expression for the phase is appropriate: $\chi=(2\pi
R_0/\lambda)\ln(R/R_0)+\chi_\odot\approx(2\pi
/\lambda)(R-R_0)+\chi_\odot$ (Mel'nik et al, 2001), what allows us
to use Fourier analysis (13).

In our coordinate system radial velocity $V_R$ is directed from
galactic centre and circular velocity deviation $\Delta
V_{\theta}$ directed along Galactic rotation. Centre of the spiral
arm corresponds to phase $\chi=0$. For definiteness (Rohlfs,
1977), we consider that the centre of the Carina--Sagittarius arm
($R\approx7$~kpc) corresponds to phase $\chi=0$, while the Perseus
arm ($R\approx9$~kpc) corresponds to phase
$\chi\approx-\frac{3}{2}\pi$.

\begin{figure}
\begin{center}
\includegraphics[width = 75mm]   {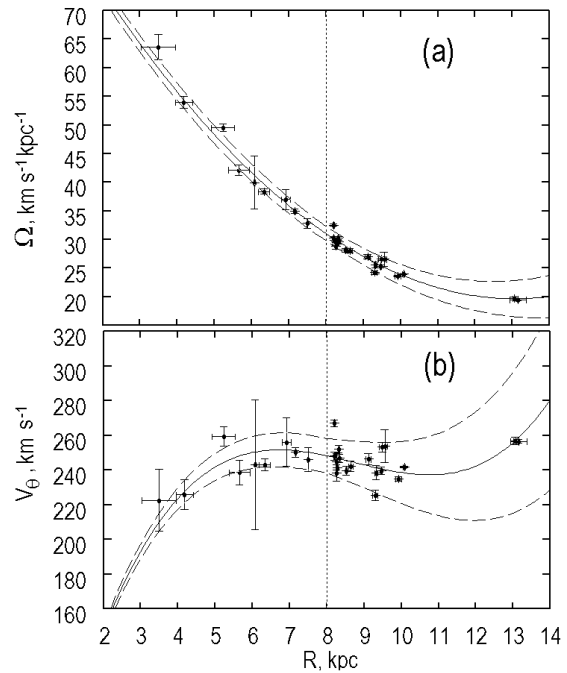}
\caption{(a)~Rotational angular velocity and (b)~rotational
circular linear velocity of the Galaxy according to (16) versus
galactocentric distance. Dashed lines denote the bounds of the
confidence interval corresponding to the $1\sigma$ level. Vertical
line indicates the location of $R_0=8$~kpc.}
\end{center}
\end{figure}

\section{RESULTS}
\subsection{Galactic Rotation Curve}

Firstly we investigated solutions of equations (1)--(3) using the
different number of Taylor expansion terms of $\Omega$ for
different values of $R_0$. Direct construction of the rotation
curve using the rotation parameters obtained above showed that
every additional expansion term leads to significant widening of
the confidence interval. We revealed that it is enough to take
three terms of Taylor expansion to construct a smooth,
sufficiently accurate Galactic rotation curve in the
galactocentric distance range of $3<R<14$~kpc. Indeed, adding
subsequent terms to the expansion practically does not change the
shape of the curve, and additional terms do not differ
statistically from zero. So, in this sense we found the optimal
solution. Results of this study are shown in Fig.~2, whence we can
see that the shape of the rotation curve significantly depends on
the adopted value of $R_0$. Thus, variation
$\mathrm{d}R_0=0.5$~kpc leads to variation of the rotational
angular velocity
$\mathrm{d}V_{\theta}\approx20\;\mathrm{km\,s}^{-1}$. As one can
see from Fig.~2, influence of inaccuracy in $R_0$ is particularly
strong in the Solar neighborhood.

Then we obtained solution of (1)--(3) for the fixed value of
$R_0=8.0$~kpc for 84 equations. As a result we found the following
parameters of the Solar velocity with respect to the group
velocity:
$(u_\odot,v_\odot,w_\odot)=(8.0,14.9,7.6)\pm(2.1,1.9,1.6)\;\mathrm{km\,
s}^{-1}$, and parameters of the angular velocity of Galactic
rotation:
$$
\displaylines{\hfill
 \Omega_0  =-31.0\pm1.2\;\mathrm{km\,s}^{-1}\mathrm{kpc}^{-1},\hfill\llap(16)\cr\hfill
 \Omega^\prime_0= +4.46\pm0.21\;\mathrm{km\,s}^{-1}\mathrm{kpc}^{-2},\hfill\cr\hfill
 \Omega^{\prime\prime}_0= -0.876\pm0.067\;\mathrm{km\,s}^{-1}\mathrm{kpc}^{-3}.\hfill
}
$$
The unit weight error is $\sigma_0=9\;\mathrm{km\,s}^{-1}$. Oort
constants
 $A=0.5R_0\Omega^\prime_0$,
 $B=\Omega_0+0.5R_0\Omega^\prime_0$
are:
 $A= 17.8\pm0.8\;\mathrm{km\,s}^{-1}\mathrm{kpc}^{-1}$ and
 $B=-13.2\pm1.5\;\mathrm{km\,s}^{-1}\mathrm{kpc}^{-1}$.
Using these values and taking into account the error
$\sigma_{R_0}=0.3$~kpc, we estimated the circular velocity of the
Solar neighborhood
$V_0=|R_0\Omega_0|=248\pm14\;\mathrm{km\,s}^{-1}$ and its orbital
period around the Galactic Centre
$T=2\pi/\Omega_0=198\pm11$~Myr.

Galactic rotational angular velocity $\Omega(R)$ and
the corresponding linear circular velocity
$V_{\theta}(R)=|R\Omega(R)|$ curves are shown in Figs.~3a and 3b,
respectively.

As one can see from Fig.~3a, errors in parameters (16) lead to
strongly increasing errors in the rotational angular velocity
$\Omega(R)$ when $R>10$~kpc. Due to this the confidence interval
for the linear rotation curve $V_{\theta}(R)$ (Fig.~3b) starts to
exceed the criterion adopted above, $\mathrm{d}V_{\theta}=20$~km
s$^{-1}$, at $R>10$~kpc. In general, we can conclude that the
Galactic rotation curve found above approximates our initial data
with fully adequate accuracy over the range of distances of
$3<R<10$~kpc.

\subsection{Periodic Perturbations}

In Fig.~4, we present distribution of deviations from circular
velocities $\Delta V_{\theta}$ versus $R$ obtained by subtraction
of the rotation curve found above, with parameters (16), from the
initial data. Fourier power spectra estimates (periodograms) of
the sequence $\Delta V_{\theta}$ ($|\overline{\Delta
V_\theta}|^2$) and galactocentric radial velocities $V_R$
($|\overline{V_R}|^2$) versus wavelength $\lambda$ are shown in
Figs.~5a and 5b, respectively. In case of $\Delta V_{\theta}$,
using two main low-frequency Fourier spectrum peaks extending up
to $\lambda=1.3$~kpc, we constructed an approximation curve of the
residuals, as shown in Fig.~4. The amplitude of the approximation
curve was fitted using a chi-square optimization technique. As it
is seen from the picture, the approximation curve is not
representative enough to make any conclusions.

\begin{figure}
\begin{center}
 \includegraphics[width = 80mm]   {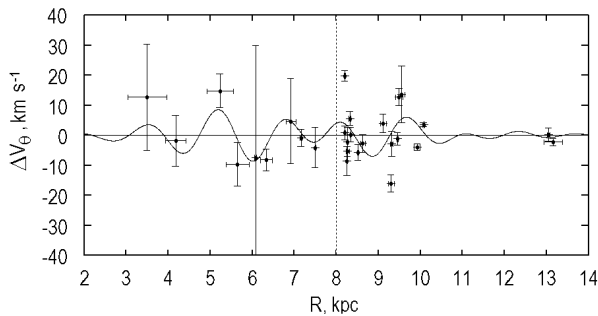}
\caption{Deviations from circular velocities $\Delta V_{\theta}$
after subtraction of the model rotation curve with parameters (16)
versus galactocentric distance; vertical line indicates
$R_0=8$~kpc.}
\end{center}
\end{figure}

\begin{figure}
\begin{center}
 \includegraphics[width = 60mm]   {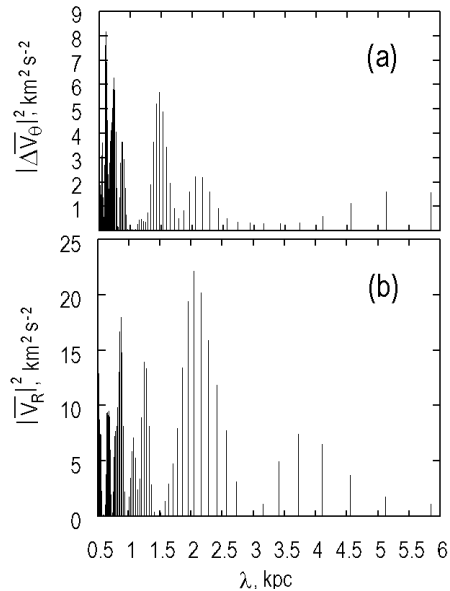}
\caption{Periodograms of (a)~deviations from circular velocities
$\Delta V_{\theta}$ and of (b)~radial velocities $V_R$.}
\end{center}
\end{figure}

But we have completely different situation in the case of radial
velocities versus $R$ shown in Fig.~6. In accordance with the main
periodogram peak shown in Fig.~5b, which is the largest one in the
frequency region of interest, perturbations in radial velocities
have an average wavelength of $\lambda=2.0\sim 0.25 R_0$~kpc,
which is in good agreement with results of other authors (see
section 5.2). The approximation curve built using this main
low-frequency Fourier spectrum peak extending from $\lambda=3$ kpc
to $\lambda=1.5$~kpc and fitted using the chi-square optimization,
is shown in bold in Fig.~6. One may clearly see that this curve
gives the most exact description of radial velocities in the
region of the Perseus arm ($R\approx9$~kpc). In this region,
extended as far as the location of the Sun, errors of amplitudes
of radial velocities are less than $\pm1\;\mathrm{km\,s}^{-1}$.
Peak value of the approximation curve reaches
$f_R=6.5\;\mathrm{km\,s}^{-1}$ and at the area close to the Sun it
is about 4.2 km s$^{-1}.$ Confidence intervals for the
approximation curve were determined using Monte-Carlo simulation
for 1000 random samples. In agreement with the approximation
curve, phase of the Sun lies in the interval
$-140^o<\chi_\odot<-120^o$ with the mean value of
$\chi_\odot\approx-130^o$. Radial perturbations in the Solar
vicinity we estimate as periodic function with wavelength
$\lambda=2\pm 0.2$ kpc and amplitude
$f_R=6.5\pm2\;\mathrm{km\,s}^{-1}$.

As it is seen from Fig.~5a and Fig.~5b both periodograms reveal
peaks at wavelength $\lambda=2$ kpc. We estimated probabilities
$p$ of belonging of these peaks to the periodic signal (not to
noise) using simple formula derived by Vityazev (2001):
$|\overline {V}(\lambda)|^2/\sigma_0^2/M\ge-\ln(1-p)$, where
$|\overline {V}(\lambda)|^2$ is periodogram peak value at
wavelength of interest, $\sigma_0^2$ is dispersion of input data
sequence, $M$ is number of data. In the first case we obtained
$p=0.6$ and in the second case $p=0.992$ what justified more
reliable estimation of spiral wave parameters from $V_R$ rather
than from $\Delta V_{\theta}$.

\subsection{Peculiar Solar Velocity with Respect to the LSR}

In Fig.~7, spatial velocities $U^1$ and $V^1$ (see Table~1), free
from Galactic rotation according to parameters (16), are given.
Residual velocity components averaged over 28 masers are
${\overline U}^1,{\overline V}^1,{\overline
W}^1=(-8.6,-13.6,-7.5)\pm(2.1,1.6,1.3)\;\mathrm{km\,s}^{-1}$, with
dispersion vector
$(\sigma_{U^1},\sigma_{V^1},\sigma_{W^1})=(10.8,8.7,6.8)\;\mathrm{km\,
s}^{-1}$.

As one may see from this figure, residual velocities of most of
masers form a sufficiently compact ellipse of smaller dispersion.
Most of deviations from this compact distribution are due to the
following three masers: G23.66$-$0.13, G23.43$-$0.18 and NGC
1333~(f2). The detail NGC 1333~(f2) reflects inner velocity
dispersion within maser NGC 1333 which itself forms an expanding
shell (Hirota et~al., 2008a). Maser G23.66$-$0.13 has
significantly negative radial velocity which reflects a certain
peculiarity in its motion.

\begin{figure}
\begin{center}
 \includegraphics[width = 80mm]   {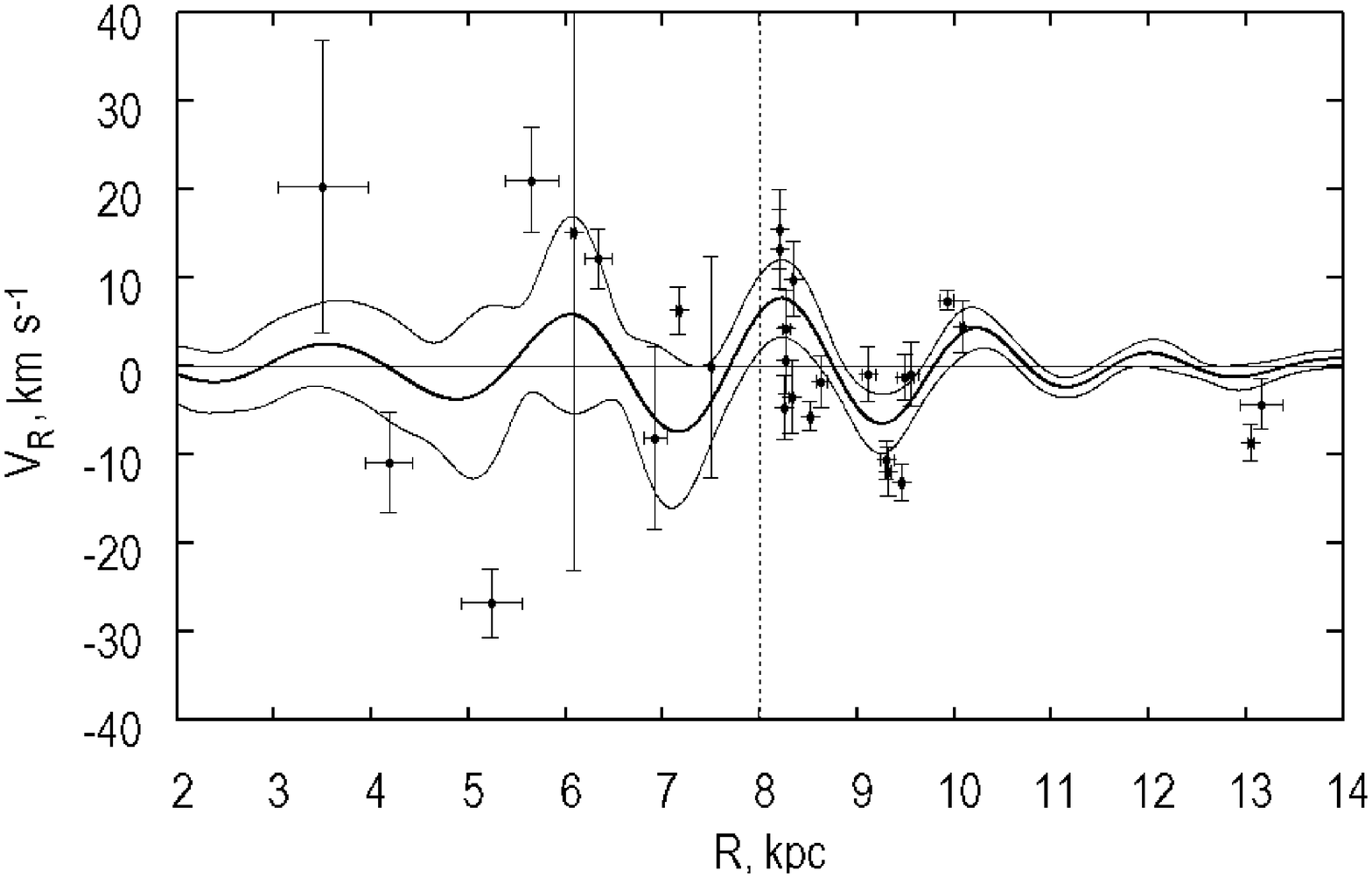}
\caption{Radial velocities of masers $V_R$ versus galactocentric
distance $R$. Thin curves mark the confidence interval
corresponding to $1\sigma$ level. Vertical line indicates the
galactocentric distance of the Sun $R_0=8$~kpc.}
\end{center}
\end{figure}

Excluding these three sources, for 25 masers forming a compact
ellipse, we deduced the following mean velocity
 ${\overline U}^1,{\overline V}^1,{\overline W}^1=(-9.1,-15.5,-8.5)\pm(1.9,1.3,1.2)\;\mathrm{km\,
s}^{-1}$, with dispersion
$(\sigma_{U^1},\sigma_{V^1},\sigma_{W^1})=(9.3,6.6,6.1)\;\mathrm{km\,
s}^{-1}$.

The question of great interest is how density wave perturbations
affect the values of the LSR velocity components. Let us denote
components of ``non-perturbed'' peculiar Solar velocity with
respect to the LSR as $(U_\odot,V_\odot,W_\odot)_\mathrm{LSR}$.
For the local case, i.e. for the immediate Solar neighborhood,
under the condition $\chi\approx\chi_\odot$ we have
$$
\displaylines{\hfill
 {\overline U}^1= -(U_\odot)_\mathrm{LSR}-V_R\cos\theta+ \Delta V_{\theta}\sin\theta,\hfill\llap(17)\cr\hfill
 {\overline V}^1= -(V_\odot)_\mathrm{LSR}+V_R\sin\theta+ \Delta V_{\theta}\cos\theta.\hfill\llap(18)
 }
$$
When $\theta\rightarrow 0~(R\approx R_0)$, we have
$\sin\theta\rightarrow 0$ and $\cos\theta\rightarrow1$. Then
$$
\displaylines{\hfill
 {\overline U}^1= -(U_\odot)_\mathrm{LSR}+      f_R\cos\chi_\odot,\hfill\llap(19)\cr\hfill
 {\overline V}^1= -(V_\odot)_\mathrm{LSR}+ f_\theta\sin\chi_\odot,\hfill\llap(20)
 }
$$
where $f_R,f_\theta>0$. As one can see from (19)--(20), signs of
corrections caused by spiral density waves are determined only by
signs of $\cos\chi_\odot$ and  $\sin\chi_\odot$. Adopting that:
mean amplitudes of spiral density wave perturbations in both
tangential and radial velocities are $6.5\pm2\;\mathrm{km\,
s}^{-1}$ and $\chi_\odot=-130^o\pm 10^o$, we obtain, according to
(19)--(20), the following ``non-perturbed'' velocity of
$(U_\odot,V_\odot,W_\odot)_\mathrm{LSR}=(5.5,11,8.5)\pm(2.2,1.7,1.2)\;\mathrm{km\,s}^{-1}$.
This result agrees with the estimate of Solar motion by
Sch\"onrich et al.~(2010) and Francis~\&~Anderson~(2009).

We could not determine the value of $\chi_\odot$ using deviations
from circular velocities of masers. But in order to confirm our
results obtained above using Galactic radial velocities $V_R$ of
masers, we analyzed also a sample of Open Star Clusters (OSCs)
from the compilation of Bobylev et al. (2007) which was based on
the catalog of Piskunov et al. (2006), along with the
line-of-sight velocities from the CRVOCA catalog (Kharchenko et
al., 2007). We selected 117 young OSCs (age $<50$~Myr) in radius
$r<2$~kpc using the following restriction for modulus of residual
velocities: $\sqrt{\Delta U^2+\Delta V^2 +\Delta
W^2}<50\;\mathrm{km\, s}^{-1}$.

Results of analysis of deviations from circular velocities $\Delta
V_{\theta}$ using Galactic rotation parameters (16) and
galactocentric radial velocities $V_R$ are shown in Figs.~8a and
8b, respectively. We can see a good agreement between the
approximation curves shown in panels (a) and (b): wavelengths are
almost similar, difference in phase is about $\pi/2$, peaks are
also comparable (6-8 km s$^{-1}$). Periodograms of deviations from
circular velocities and radial velocities  are shown in Fig.~9. To
obtain approximation curves in Fig.~8, we used low-frequency
Fourier spectra up to wavelengths $\lambda=1.3$~kpc and fitted the
amplitude of perturbations using the chi-square optimization.
Spectral analysis of $V_R$ (Fig.~9b) reveals periodic motion with
wavelength $\lambda=1.9 (0.23R_0)$~kpc. Spectral analysis of
$\Delta V_{\theta}$ (Fig.~9a) displays three dominant peaks at
wavelengths $\lambda=2.21 (0.28R_0), 1.1 (0.13R_0), 0.47
(0.058R_0)$~kpc (compare with results by Clemens
(1985):$\lambda=0.22R_0, 0.12 R_0, 0.05 R_0$).

Now we can compare these results with those obtained using radial
velocities $V_R$ of masers. We can see a good agreement in the
behavior of the approximation curves shown in Fig.~8b and Fig.~6
in the Solar area. Fourier spectrum of radial velocities of masers
$V_R$ reveals a dominant peak of amplitude $6.5\;\mathrm{km\,
s}^{-1}$  at wavelength $2$~kpc. In addition, there is a good
agreement of phase $\chi_\odot\approx -120^o$ using data shown in
Fig.~8b with the result obtained from masers $\chi_\odot\approx
-130^o\pm 10^o$.

\section{DISCUSSION}
\subsection{Galactic Rotation Curve}

The shape of the Galactic rotation curve obtained above is in a
good agreement with the one obtained by Zabolotskikh et al. (2002)
for $R_0=7.5$~kpc as a result of thorough reconciliation of
distance scales of different star groups. The value
$\Omega_0=-31.0\pm1.2\;\mathrm{km\,s}^{-1}\mathrm{kpc}^{-1}$ (see
solution (16)) is in a good agreement with the following result
obtained from 18 masers: $-29.9\;\mathrm{km\,
s}^{-1}\mathrm{kpc}^{-1}<\Omega_0<-31.6\;\mathrm{km\,s}^{-1}\mathrm{kpc}^{-1}$
(McMillan~\&~Binney, 2010) and with
$\Omega_0=-30.3\pm0.9\;\mathrm{km\,s}^{-1}\mathrm{kpc}^{-1}$ (Reid
et al., 2009). Our value $V_0=248\pm14\;\mathrm{km\,s}^{-1}$
($R_0=8$ kpc) is in a good agreement with
$254\pm16\;\mathrm{km\,s}^{-1}$  ($R_0=8.4$ kpc) by Reid et al.
(2009) and $244\pm13\;\mathrm{km\,s}^{-1}$ ($R_0=8.2$ kpc) by Bovy
et al. (2009).

\subsection{Spiral Structure Parameters}

The estimate $\lambda=0.22R_0$ of the distance between adjacent
spiral arms was obtained by Clemens (1985). For $R_0=8.5$~kpc used
here it gives $\lambda=1.8$~kpc which is in a good agreement with
the value $\lambda=2.0$~kpc obtained in our study using 28
selected masers, as well as with $\lambda=1.9$ obtained from 117
young OSCs. Mel'nik et al.~(2001) obtained  the estimate
$\lambda=2.0\pm0.2$~kpc by analyzing velocities of OB associations
in the 3-kpc Solar neighborhood. An independent estimate
$\lambda=1.7\pm0.5$~kpc was obtained from the analysis of $V_R$
velocities of young OSCs in the distance range of $6<R<9$~kpc
(Bobylev et al., 2008).

From the analysis of OB stars and Cepheids,
Byl~\&~Ovenden (1978) found the following parameters of the
Galactic spiral structure
 $f_R=3.6\pm0.4\;\mathrm{km\,s}^{-1}$ and
 $f_\theta=4.7\pm0.6\;\mathrm{km\,s}^{-1}$, $\chi_\odot\approx165\pm1^\circ$, $i=-4.2\pm0.2^\circ$.
According to Clemens~(1985), the value of perturbation amplitude
is $|f_\theta|=5\;\mathrm{km\,s}^{-1}$ at {\bf the} wavelength of
$\lambda=0.22R_0$.

\begin{figure}
\begin{center}
\includegraphics[width = 55mm]   {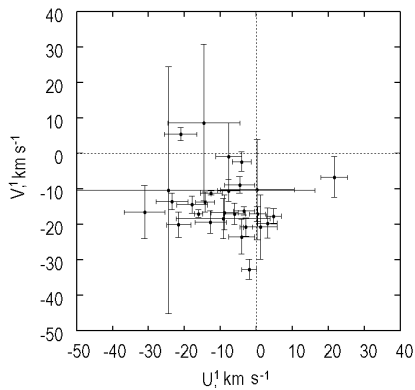}
\caption{Heliocentric spatial velocities  $U^1$ and $V^1$ free
from Galactic rotation with parameters~(16).}
\end{center}
\end{figure}

Using Cepheids, Mishurov et al.~(1997) found the following
parameters:
 $f_R=6.3\pm2.4\;\mathrm{km\,s}^{-1}$ and
 $f_\theta=4.4\pm2.4\;\mathrm{km\,s}^{-1}$, $\chi_\odot\approx290\pm16^\circ$, $i=-6.8\pm0.7^\circ$.
Using Hipparcos OB stars and Cepheids, Fern\'andez et al.~(2001)
showed that amplitudes $f_R$ and $f_\theta$ are about
$6-4\;\mathrm{km\, s}^{-1}$ and  $20^\circ<\chi_\odot<284^\circ$.
Mean amplitude of radial perturbations
$f_R=6.5\pm2\;\mathrm{km\,s}^{-1}$ found by us using masers is in
a good agreement with the estimates listed above.

Location of the Sun in the external part of the
Carina--Sagittarius arm is in agreement both with the distribution
of stars and gas (Russeil, 2003) and with density wave theory.

\subsection{Peculiar Velocity of the Sun}

It is worth mentioning some results of determination of the non-perturbed LSR
velocity parameters obtained simultaneously with
parameters of the Galactic rotation, taking into account
perturbations evoked by the spiral wave density. For example,
these are: $(U_\odot,V_\odot)_\mathrm{LSR}=(2,5)\pm(2,2)\;\mathrm{km\,s}^{-1}$
(Mishurov et al., 1997);
$(U_\odot,V_\odot)_\mathrm{LSR}=(7.8,13.6)\pm(1.3,1.4)\;\mathrm{km\,s}^{-1}$
(Mishurov~\&~Zenina, 1999);
$(U_\odot,V_\odot)_\mathrm{LSR}=(9,12)\pm(1,1)\;\mathrm{km\,s}^{-1}$ (L\'epine et
al.,~2001). Let us mention also the results of Fern\'andes et al.~(2001),
obtained using different samples of OB stars:
$(U_\odot,V_\odot,W_\odot)_\mathrm{LSR}=(8.8,12.4,8.4)\pm(0.7,1.0,0.5)\;\mathrm{km\,
s}^{-1}$ and of Cepheids:
$(U_\odot,V_\odot,W_\odot)_\mathrm{LSR}=(6.5,10.4,5.7)\pm(1.2,1.9,0.7)\;\mathrm{km\,
s}^{-1}$.

From the results listed here, we can see that values of ($U,V$)
components of the LSR velocity are always smaller than the
corresponding components of the standard Solar motion velocity
$(U_\odot,V_\odot,W_\odot)=(10.3,15.3,7.7)\;\mathrm{km\,s}^{-1}$.

\begin{figure}
\begin{center}
\includegraphics[width = 60mm]   {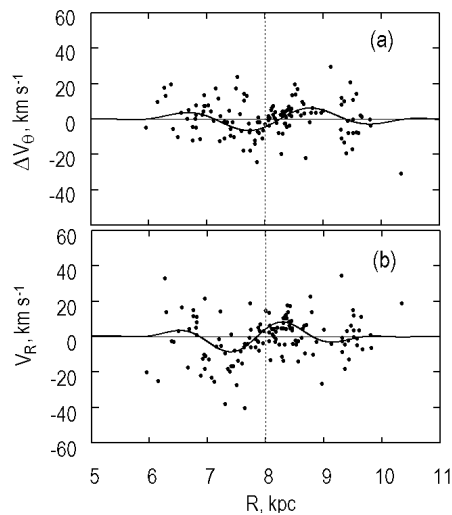}
\caption{Deviations from (a)~circular velocities $\Delta
V_{\theta}$ and (b)~radial velocities $V_R$ for 117 young OSCs
versus galactocentric distance $R$. Approximation curves are shown
with solid lines. Vertical line indicates galactocentric distance
of the Sun $R_0=8$~kpc. }
\end{center}
\end{figure}

\begin{figure}
\begin{center}
\includegraphics[width = 60mm]   {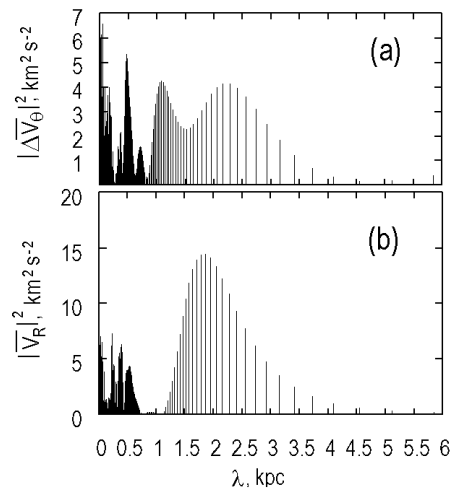}
\caption{Periodograms of deviations from (a)~circular velocities
$\Delta V_{\theta}$ and (b)~radial velocities $V_R$ for 117 young
OSCs.}
\end{center}
\end{figure}

\section{CONCLUSIONS}

Spatial velocities of all currently known 28 masers having
estimates of trigonometric parallaxes are reanalyzed. To determine
the Galactic rotation curve, we used the first three terms of
Taylor expansion of the Galactic angular rotational velocity at
the galactocentric distance of the Sun $R_0=8.0$~kpc. Coefficients
obtained from Bottlinger's equations are:
 $\Omega_0  =-31.0\pm1.2\;\mathrm{km\,s}^{-1}\mathrm{kpc}^{-1}$,
 ${\Omega_0^{'}}= 4.46\pm0.21\;\mathrm{km\,s}^{-1}\mathrm{kpc}^{-2}$,
 $\Omega_0^{''}=-0.876\pm0.067\;\mathrm{km\,s}^{-1}\mathrm{kpc}^{-3}$.
Oort constants are
 $A=17.8\pm0.8\;\mathrm{km\,s}^{-1}\mathrm{kpc}^{-1}$ and
 $B=-13.2\pm1.5\;\mathrm{km\,s}^{-1}\mathrm{kpc}^{-1}$, which allowed
us to estimate the circular velocity of the Solar neighborhood
rotation $V_0=248\pm14~\mathrm{km\,s}^{-1}$ ($R_0=8$ kpc).

It was found that deviations from circular velocities $\Delta
V_{\theta}$ of masers are not representative enough for any
reliable analysis, while the study of galactocentric radial
velocities $V_R$ of masers allowed us to estimate a number of
useful parameters concerning spiral structure of the Galaxy.
Based on Fourier analysis of velocities $V_R$, we obtained that:
there are periodic perturbations evoked by the spiral density
wave, with wavelength of $\lambda=2.0\pm0.2$~kpc and peak velocity
of $f_R=6.5\pm2\;\mathrm{km\,s}^{-1}$ and velocity of
$4\pm1\;\mathrm{km\,s}^{-1}$ near the location of the Sun; phase
of the Sun in the density wave is $\chi_\odot=-130^o\pm10^o$,
which proves that the Sun is located in the inter-arm space.

Adopting the parameters of spiral density wave found we determined
the following non-perturbed Solar peculiar velocity with respect
to the LSR:
$(U_\odot,V_\odot,W_\odot)_\mathrm{LSR}=(5.5,11,8.5)\pm(2.2,1.7,1.2)\;\mathrm{km\,s}^{-1}$.
In this connection, in the regions near the Sun, velocity
component $(V_\odot)_\mathrm{LSR}$ is influenced mostly by
negative perturbation $\Delta V_{\theta}$ and has smaller value
than the corresponding velocity component
$V_\odot=15.3\;\mathrm{km\, s}^{-1}$ of the Standard Solar motion.

\section*{ACKNOWLEDGMENTS}

The authors are thankful to the anonymous referee for critical
remarks which promoted improving the paper. This study was
supported by the Russian Foundation for Basic Research (project
No. 08--02--00400), and in part by the ``Origin and Evolution of
Stars and Galaxies'' --- ``Program of the Presidium of the Russian
Academy of Sciences and the Program for State Support of Leading
Scientific Schools of Russia'' (NSh--6110.2008.2). The authors are
thankful to Vladimir Kouprianov for his assistance in preparing
the text of manuscript.

{
}
\end{document}